\documentstyle[epsfig,twocolumn]{esapub}

\newcommand{\siiv}{Si~{\sc iv}}
\newcommand{\civ}{C~{\sc iv}}
\newcommand{\nv}{N~{\sc v}}
\newcommand{\kms}{km~s$^{-1}$}
\newcommand{\vinf}{$v_{\infty}$}
\newcommand{\Ha}{H$\alpha$}
\newcommand{\0}{\verb+ +}

\begin{document}

\setlength{\parindent}{0pt}
\setlength{\parskip}{ 10pt plus 1pt minus 1pt}
\setlength{\hoffset}{-1.5truecm}
\setlength{\textwidth}{ 17.1truecm }
\setlength{\columnsep}{1truecm }
\setlength{\columnseprule}{0pt}
\setlength{\headheight}{12pt}
\setlength{\headsep}{20pt}
\pagestyle{esapubheadings}

\title{\bf WINDS IN OB-TYPE STARS}

\author{{\bf Lex~Kaper} \vspace{2mm} \\
European Southern Observatory, Karl-Schwarzschil-Str. 2, D-85748
Garching bei M\"{u}nchen, Germany}

\maketitle

\begin{abstract}
The International Ultraviolet Explorer satellite has made a tremendous
contribution to the study of hot-star winds. Its long lifetime has
resulted in the collection of ultraviolet spectra for a large sample
of OB~stars. Its unique monitoring capability has enabled detailed
time-series analyses to investigate the stellar-wind variability for
individual objects. IUE has also been a major driver for the
development of the radiation-driven-wind theory; the synergy between
theory and observations is one of the main reasons for the large
progress that has been made in our understanding of hot-star winds and
their impact on the atmospheres and evolution of massive stars.
\vspace {5pt} \\

  Key~words: hot stars; stellar winds; variability.
\end{abstract}

\section{INTRODUCTION}

OB-type stars, with masses ranging from about 8 to up to more than 100
M$_{\odot}$, are hot ($>10,000$~K) and luminous
($10^{5}-10^{6}$~L$_{\odot}$). The brightest supergiants with an
absolute visual magnitude M$_{V} = -9$ are visible up to the Virgo
cluster (m$_{V} = 22$) and thus potential candidates for the study
of stellar populations in external galaxies. Because of their high
luminosity, OB-type stars do not grow old (lifetime $\sim 1-100 \times
10^{6}$ years) and thus trace star-forming regions.

From the shape of several strong spectral lines in their spectrum
(like, e.g., the ultraviolet resonance lines of \nv, \siiv, and \civ),
it becomes immediately clear that OB~stars lose a vast amount of
material through a stellar wind. The blue edges of these so-called
P~Cygni-type profiles indicate maximum outflow velocities (few
$100-4000$ km/s) exceeding the surface escape velocity by a factor
between 1 and 3 (Abbott \cite{Ab78}, Prinja et al.\ \cite{PB90}). The
associated mass-loss rates, most reliably measured from the stellar
wind's free-free emission observed at radio wavelengths (Bieging et
al. 1989), strongly depend on the OB-star's luminosity
(\.{M}$\propto$L$^{1.6}$) and range from about $10^{-9}$ to more than
$10^{-5}$ M$_{\odot}$/yr. Thus, during their lifetime OB~stars lose a
significant fraction of their initial mass, making mass loss a key
ingredient in the description of their evolution (cf.\ Maeder \& Conti
\cite{MC94}, Chiosi \cite{Ch98}). Furthermore, hot-star winds play a
dominant role in providing momentum, energy, and nuclearly processed
material to the interstellar medium (Abbott \cite{Ab82}, Castor
\cite{Ca93}).

\begin{figure*}[!ht]
\centerline{
\hbox{\epsfig{file=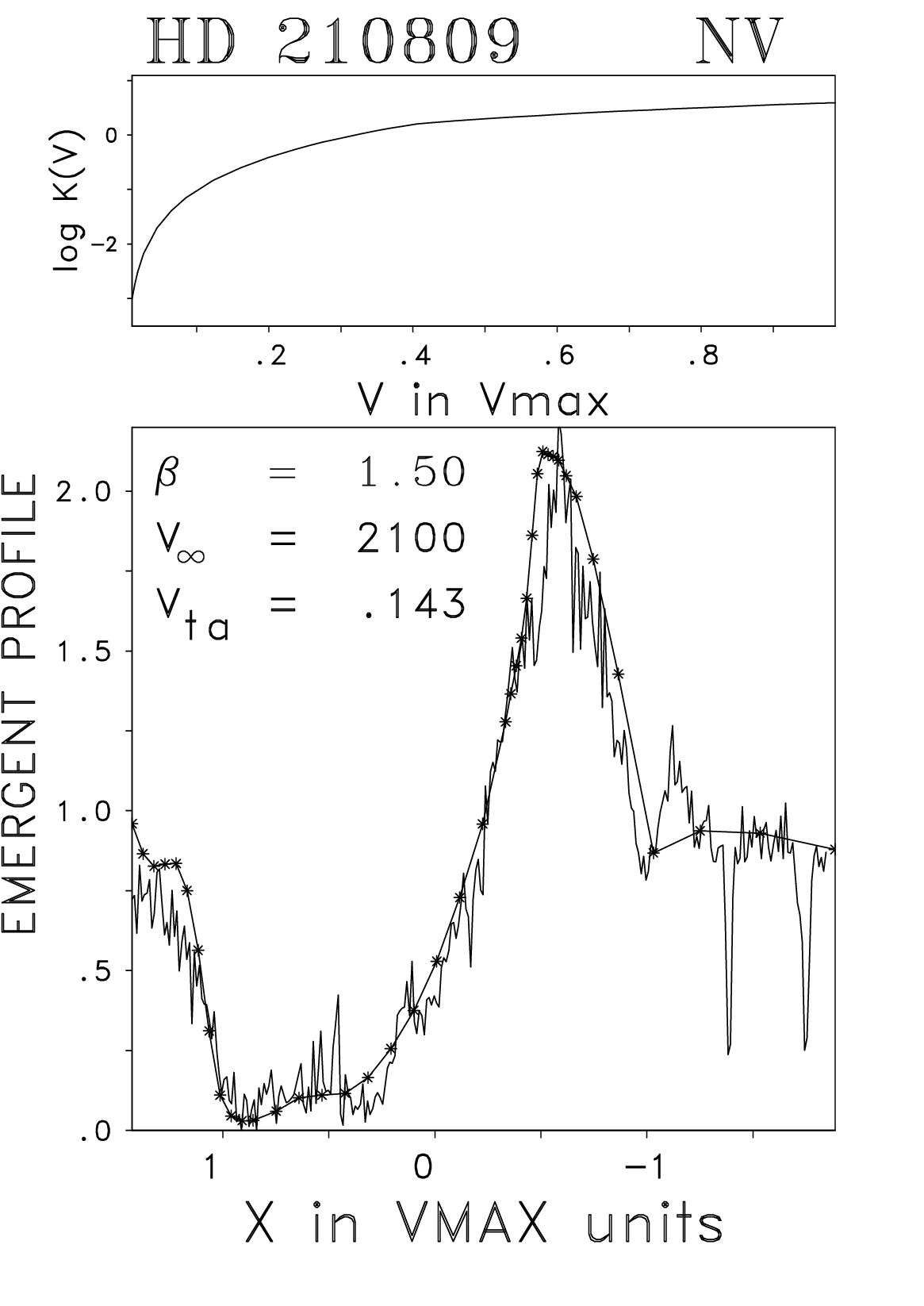,width=5cm}
      \epsfig{file=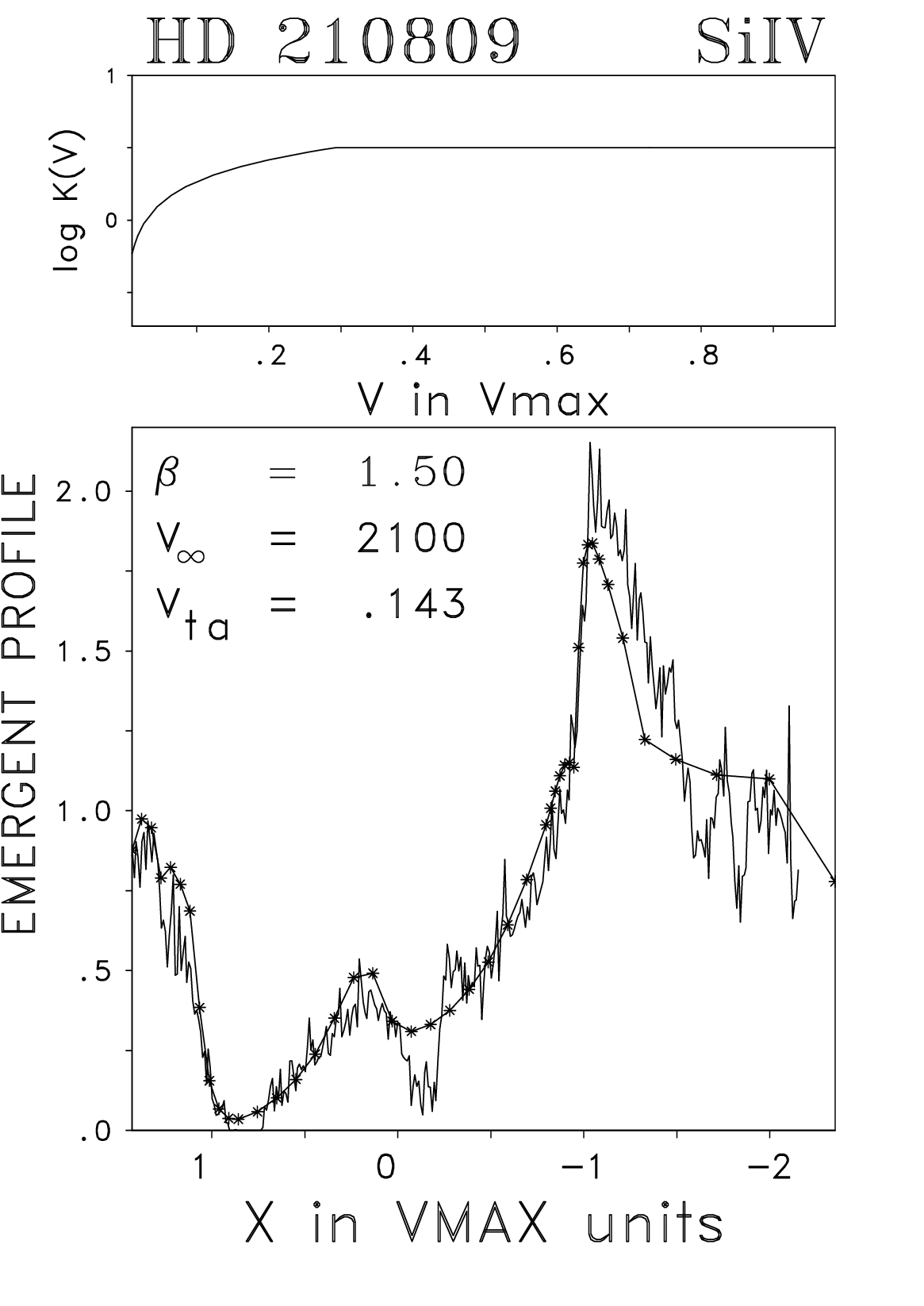,width=5cm}
      \epsfig{file=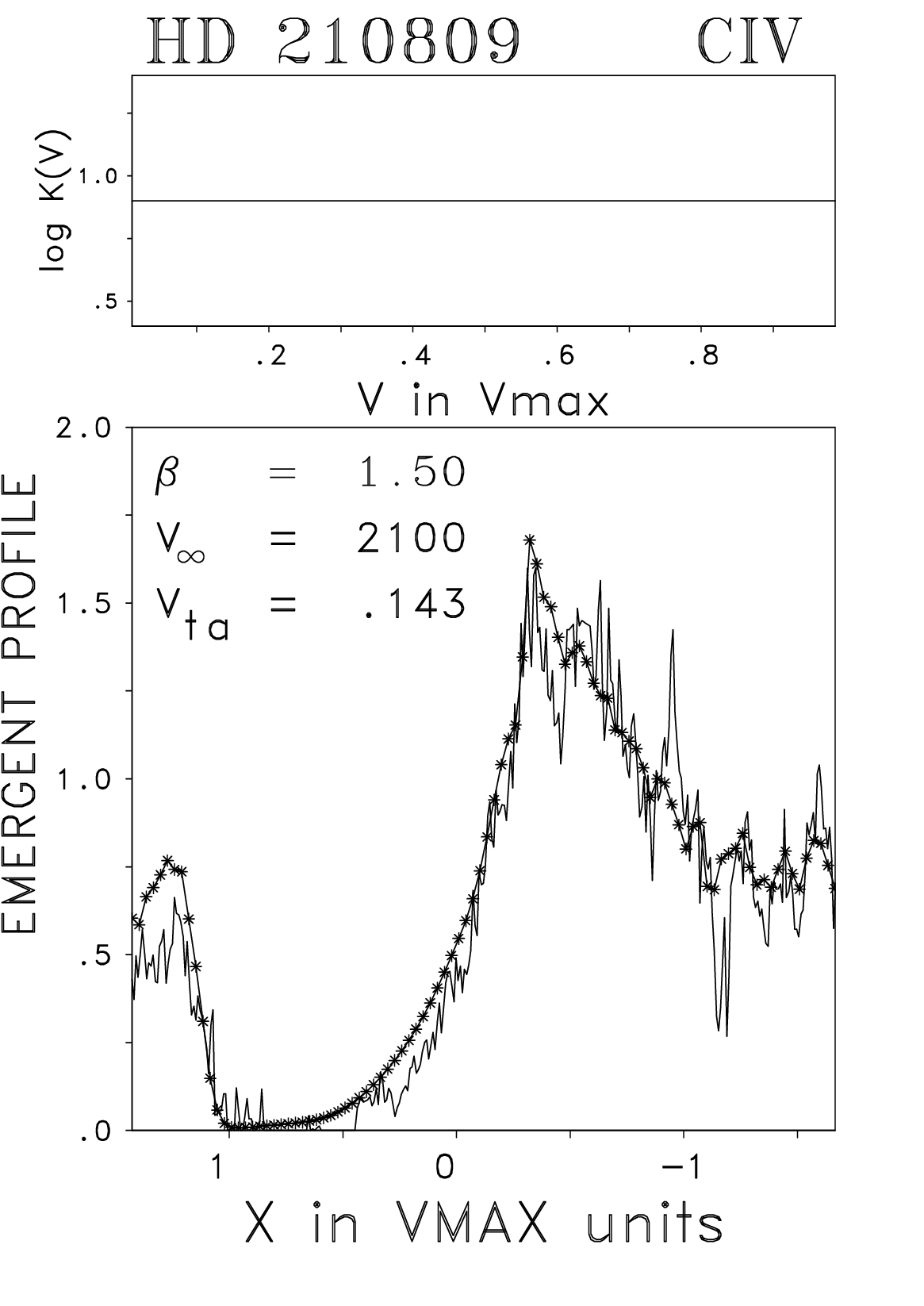,width=5cm}}}
\caption[]{\em Radiative transfer fits of the ultraviolet (IUE)
resonance lines of \nv, \siiv, and \civ\ of the galactic O9~Iab
supergiant HD210809. The x-axis is the velocity displacement measured
from line center (blue doublet component) in units of \vinf. The
corresponding values for $\beta$, \vinf, and turbulent velocity
$v_{ta}$ are listed. The upper panels display the fit of the line
strength stratification $k(v) \propto \epsilon_{\rm abund} X_{\rm
ion}$\.{M} as a function of velocity (from Haser \cite{Ha95}).}
\end{figure*}

In the following we will concentrate on the contribution of IUE to the
study of hot-star winds. Why was the IUE satellite so important for
this field of research?  First of all, the main part of the OB-star
spectrum is emitted in the far-UV, and, due to atomic physics, that is
also the region where most of the important lines are situated. The UV
resonance lines provide excellent diagnostics of the stellar-wind
structure. Knowledge of the wind structure is essential if one wants
to derive the stellar atmospheric parameters (cf.\ Kudritzki
\cite{Ku88}), since practically all spectral lines are affected by the
presence of a stellar wind. An important point we want to stress in
this paper is that the success of IUE in this field of astronomy is
for a substantial part based on the rapid and parallel progress made
by theory. Each time a new IUE proposal had to be written we could
report on the recent theoretical developments and predictions which
could be tested with new observations.

The IUE archive contains high-resolution spectra ($R=10,000$) for a
large sample of OB~stars ($\sim 400$), which has enabled the study of
stellar-wind properties as a function of spectral type. Several
atlases have been constructed showing the IUE spectra of O stars
(Walborn et al.\ \cite{WN85}), B stars (Rountree \& Sonneborn
\cite{RS93}), and the UV P~Cygni profiles of stars with spectral types
between O3 and F8 (Snow et al.\ \cite{SL94}). The ``zoo'' of OB stars
contains many different species: ``normal'' OB main sequence stars,
giants, supergiants; subdwarf OB stars (Heber); Wolf-Rayet stars
(Willis); Luminous Blue Variables (Shore, Wolf); Central Stars of
Planetary Nebulae; Be stars (Smith); B[e] stars; $\beta$~Cep stars;
magnetic B stars; hot binaries (Stickland); etcetera. Within brackets
reference is made to the author of a contribution in these proceedings
covering the respective type of stars; we will focus on the ``normal''
OB stars.

The second part of this paper deals with the variability aspect of
hot-star winds, a topic that could be studied because of the unique
monitoring capacity of IUE. Mainly on the basis of IUE observations it
could be demonstrated that variability is a fundamental property of
radiation-driven winds, that it is not chaotic, but systematic, and
that its characteristic timescale is determined by the stellar
rotation period.

\section{THE PHYSICS OF HOT-STAR WINDS}

That the driving of hot-star winds is based on the interplay between
gravity and the stellar radiation field becomes already apparent from
the observed relation between the terminal velocity of the wind,
\vinf, and the escape velocity (Abbott \cite{Ab78}), and the relation
between the mass-loss rate and stellar luminosity (Garmany \& Conti
\cite{GC84}). In principle, one should be able to inverse the latter
relation in order to derive the luminosity, and thus the distance,
from the observed mass-loss rate, were it not for the relatively large
scatter in this relation.

Radiation-driven winds work on the principle that momentum contained
in the stellar radiation field is transferred to gas particles in the
wind via the scattering of photons. The main point is that momentum is
a vector quantity and that the photons before scattering are all
moving in one direction, i.e. away from the star, while they move in a
``random'' direction after the first scattering. The result is that
the associated radiative force is directed away from the star. The
scattering process takes place via spectral lines of the gas particles
in the outer atmosphere. Lucy \& Solomon (\cite{LS70}) were among the
first to realize that the scattering of photons over a few strong
(resonance) lines would result in a strong enough force to drive a
wind. An essential ingredient is that the wind on its way out reaches
velocities which are about a 100 times larger than the typical thermal
width of the spectral lines, so that due to the doppler shift of the
lines many more photons can be ``tapped'' compared to the static
case. Castor, Abbott, and Klein (CAK, \cite{CA75}) improved upon this
result by incorporating also weak stellar lines, and were able to
qualitatively reproduce the observed relations between \vinf\ and
$v_{{\rm esc}}$, and between \.{M} and L.

\begin{figure}[ht]
\centerline{\epsfig{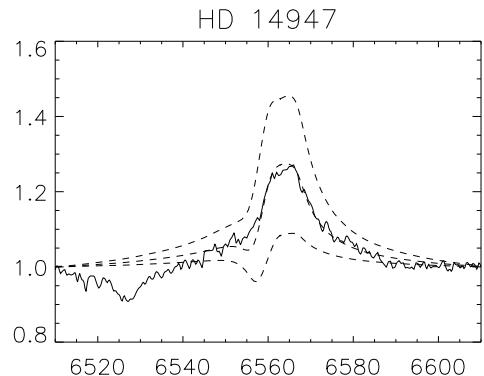}}
\caption[]{\em \Ha\ profile (drawn line) of the O5~Iaf+ supergiant
HD14947 compared with unified model calculations adopting $10, 7.5$,
and $5.0 \times 10^{-6}$ M$_{\odot}$/yr, respectively, for the
mass-loss rate (Puls et al.\ \cite{PK96}, figure from Kudritzki
\cite{Ku98}).}
\end{figure}

This was the state of the theory in the pre-IUE era. After the
collection of ultraviolet spectra of many OB~stars and measuring
\vinf\ and \.{M} for a large sample of stars, it became clear that CAK
theory had to be modified to quantitatively explain the
observations. Independently, Friend \& Abbott (\cite{FA86}) and
Pauldrach, Puls, and Kudritzki (\cite{PP86}) improved on CAK theory by
dropping some of the assumptions, the most important one being that
the star is extended instead of a point source. Since then,
radiation-driven wind theory has been further refined, in parallel
with the development of non-LTE atmosphere models for hot stars; more
than a million of different lines from over a hundred different ionic
species are taken into account to calculate the radiative force (e.g.\
Pauldrach et al.\ \cite{PK94}) and the resulting hydrodynamic
structure of the stellar wind. It is now possible to model OB-star
spectra quantitatively, consistently treating the photospheric and
wind lines using so-called unified model atmospheres (cf.\ Kudritzki
\& Hummer \cite{KH90}, Kudritzki \cite{Ku98}).

Figure 1 shows model fits to the UV resonance lines of the O9~Iab
supergiant HD210809 (Haser \cite{Ha95}). An improvement with respect
to the SEI method (Lamers et al.\ \cite{LC87}, Groenewegen \& Lamers
\cite{GL89}) is that Haser's method allows a dependence of the
turbulent velocity $v_{ta}$ on velocity, which is predicted by theory
(see below). The fits can be used to determine \.{M}, \vinf, and the
$\beta$-parameter (which defines the steepness of the velocity
law). Although \vinf\ can be derived very accurately (typically 5\%
accuracy), the determination of \.{M} is very difficult, mainly
because of the large uncertainty in the degree of ionization $X_{\rm
ion}$. Instead, the mass-loss rate can be much more precisely obtained
from the \Ha\ profile (Puls et al.\ \cite{PK96}). An example is given in
Figure 2 for the O5~Iaf+ supergiant HD14947. A nice demonstration of
the present state of the art is presented by Taresch et al.\ (1997),
who model the most luminous and most massive star in our galaxy,
HD93129A. One of their conclusions is that modelling of the far-UV
spectrum is only possible when taking the interstellar absorption
spectrum properly into account.

\begin{figure}[ht]
\centerline{\epsfig{file=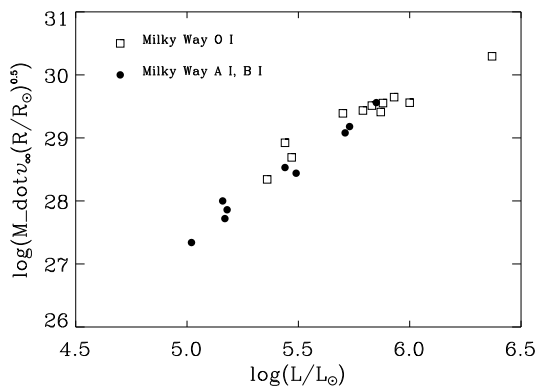,width=8.0cm}}
\caption[]{\em The observed Wind Momentum - Luminosity Relation for
A-, B-, and O-supergiants in the Galaxy. All mass-loss rates have been
determined from \Ha; the terminal velocities were derived from the
blue edges of UV P~Cygni profiles (OB supergiants) or from \Ha\ (A
supergiants). The figure is taken from Kudritzki (\cite{Ku98}).}
\end{figure}

\subsection{The Wind Momentum Luminosity Relation}

As discussed in the previous section, the physics of hot-star winds
strongly depend on the stellar radiation field. Would it be possible
to determine the OB-star luminosity from its stellar-wind properties?
In a recent review, Kudritzki (\cite{Ku98}) argues that the observed
relation between wind momentum (\.{M}\vinf) and luminosity is very
tight and can be used to derive distances to OB~stars (see also Lamers
\& Leitherer \cite{LL93}, Puls et al.\ \cite{PK96}). Leaving out some
terms with only a weak dependence on (other) stellar parameters, one
obtains from theory a wind-momentum luminosity (WML) relation of the
following form (Kudritzki \cite{Ku98}):
\[ \mbox{\.{M}}v_{\infty} \propto \frac{1}{\sqrt{R_{\star}}} L^{1/\alpha}, \]
where $\alpha \approx 2/3$ is a parameter which depends on the shape
of the line strength distribution and the ionization structure of the
stellar wind. This already indicates that the WML relation should
depend on metallicity. 

Using the techniques described above, an empirical WML relation can be
constructed by deriving \.{M} and \vinf\ from the spectra of a number
of stars with known luminosity (and radius!). The result based on a
sample of galactic supergiants of type A, B, and O is shown in Figure
3 (Puls et al.\ \cite{PK96}, Kudritzki \cite{Ku98}). Indeed, a tight
WML relation is observed, with possibly a slightly different slope for
the O supergiants with respect to the A and B supergiants (filled
circles). The empirical WML relation can be calibrated as a function
of metallicity by observing blue supergiants in nearby galaxies with a
different metallicity (e.g. LMC, SMC). When this has been done, the
WML relation can be used to determine distances to blue supergiants
beyond the local group, out to the Virgo and Fornax clusters of
galaxies. Kudritzki (\cite{Ku98}) argues that the uncertainties in
distance modulus obtained with the WML method are comparable to those
for individual Cepheids in galaxies if the period-luminosity relation
is applied. The advantage of the WML method is that an individual
reddening (and therefore extinction) can be derived for every object
by comparing its observed colors with the model predictions. However,
the possible impact of spectral variability (see section 3) on these
results has to be investigated.

\subsection{Instability of Radiation-Driven Winds}

Radiation-driven winds are excellent laboratories for the study of
radiation hydrodynamics; fundamental contributions to this field of
physics stem from the observations of hot-star winds. Hydrodynamical,
time-dependent (1-dimensional) simulations of radiation-driven winds
have shown that the acceleration mechanism contains a potent
instability. In their seminal paper, Lucy \& Solomon (\cite{LS70})
already remarked that radiation-driven flows are unlikely to be
steady. A small perturbation in velocity of a fluid element with
respect to the surrounding gas will result in a strong increase (or
decrease) of the acceleration of the ions in this element, because
they are suddenly exposed to the unattenuated stellar continuum flux
(or ``shadowed'' by the low-velocity wind closer to the star). This
instability leads to the formation of shocks caused by high-velocity,
low-density material running into low-velocity, high-density regions
in front (Owocki et al. \cite{OC88}, Owocki \cite{Ow92}). Thus, a
radiation-driven wind is expected to be highly structured on
relatively small scales.

The non-monotonic velocity and density structure of radiation-driven
winds and the predicted formation of shocks provide a solution to a
number of important observational facts: (i) the saturation of strong
UV resonance lines; (ii) the observed soft X-ray emission of hot
stars; and (iii) the observation of ions (e.g. O~{\sc vi}, \nv) in the
wind that cannot be explained by the temperature of the radiation
field (so-called super ionization). Lucy (\cite{Lu82a}) showed that in
a shocked stellar wind a photon has to penetrate several (in stead of
one) ``resonance zones'' before it can escape. This explains the
formation of P~Cygni profiles with black absorption troughs (as in
Figure~1, see Puls et al.\ \cite{PO93}). The observed soft X-ray
emission from hot stars (Chlebowski et al.\ \cite{CH89}, Bergh\"{o}fer
et al.\ \cite{BS97}) cannot be due to the star itself or a hypothetical
corona since the stellar wind is optically thick at soft X-ray
wavelengths. The X-ray emission has to originate in the outer wind
regions. The existence (and interaction) of shocks in the wind could
provide a natural explanation (Lucy \cite{Lu82b}, Feldmeier
\cite{Fe95}). The super ionization might as well be the result of the
produced X-ray flux in the wind (Pauldrach et al.\ \cite{PK94}).

\begin{figure}[!t]
\centerline{\epsfig{file=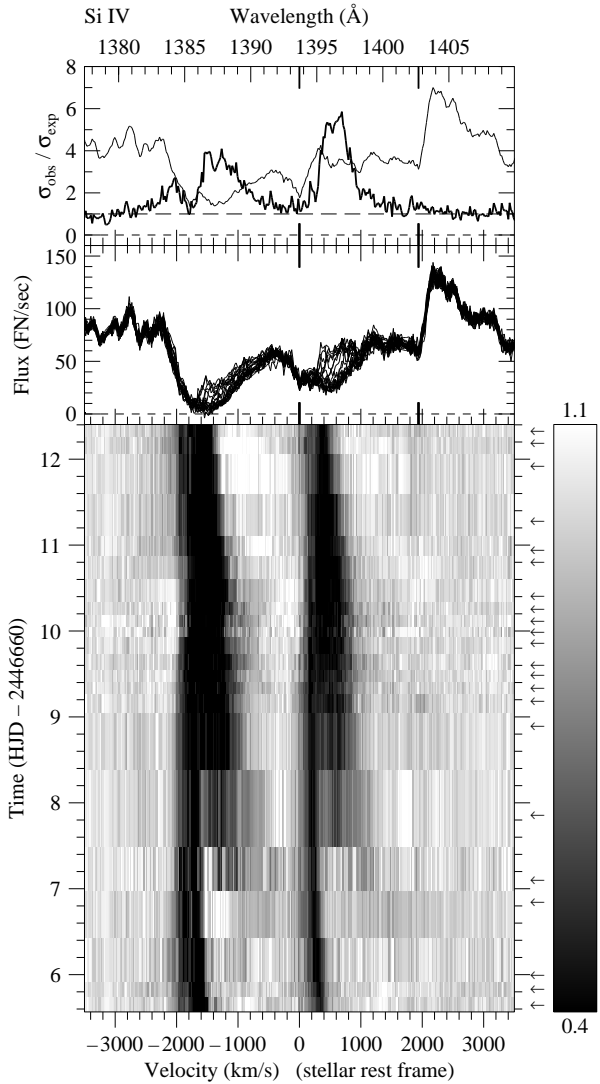,width=8.0cm}}
\caption[]{\em 19~Cep O9.5Ib ($v \sin{i}=75$ km/s): Time series of the
\siiv\ resonance doublet obtained in August 1986. The top panel
displays the minimum-absorption template (thin line) used to produce
the residual spectra shown in the grey-scale figure. The level of
variability is indicated by a thick line. The middle panel shows an
overplot of the spectral lines. In the lower panel the residual fluxes
are converted into levels of grey; arrows denote mid-exposure times. A
DAC appears at day 7 in both components of the doublet and slowly
accelerates to its asymptotical velocity (from Kaper et al.\
\cite{KH96}).}
\end{figure}

\section{VARIABILITY IN HOT-STAR WINDS}

\begin{figure}[ht]
\centerline{\epsfig{file=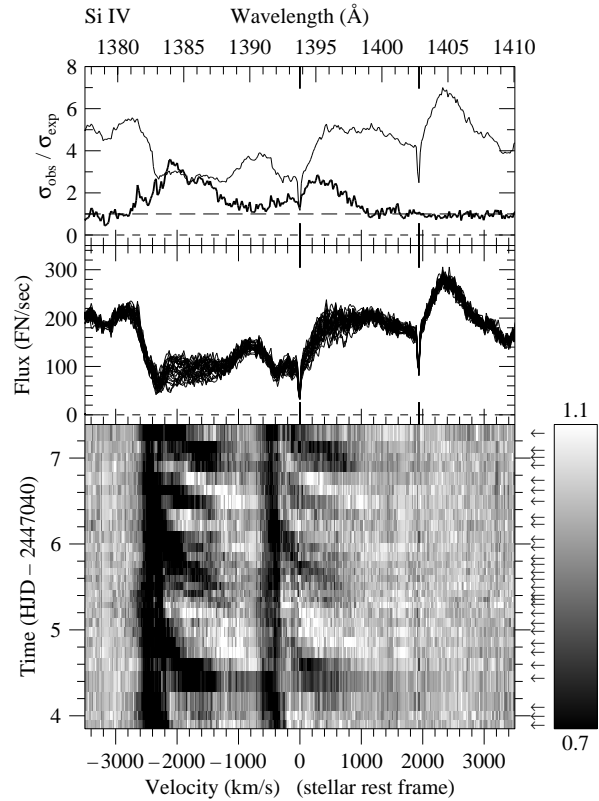,width=8.0cm}}
\caption[]{\em 68 Cyg O7.5 III:n((f)) ($v \sin{i}=274$ km/s): Time
series \siiv\ resonance doublet (as Figure~4) obtained in September
1987. In both doublet components the appearance and evolution of 4 DAC
events is witnessed. The absorption components reach an asymptotic
velocity of 2400 km/s, i.e. \vinf. The recurrence and acceleration
timescale of the DACs is much shorter (1.4~day) compared to 19~Cep
in Figure~4 (from Kaper et al.\ \cite{KH96}).}
\end{figure}

IUE has proven to be a powerful tool for the study of variability in
the supersonically expanding winds of hot stars. In particular, the
blue-shifted absorption parts of the UV P~Cygni lines show dramatic
changes with time. Some of these profiles contain ``narrow''
absorption components, first recognized in {\it Copernicaus} spectra
of OB-type stars, at velocities close to the terminal velocity of the
wind (e.g.\ Underhill \cite{Un75}, Snow \& Morton \cite{SM76}). Lamers
et al.\ (\cite{LG82}) reported the presence of narrow components in 17
out of 26 OB stars at a typical velocity of 0.75~\vinf\ and a mean
width of about 0.18~\vinf. Later time-resolved studies with IUE (e.g.\
Henrichs \cite{He84}, Prinja \& Howarth \cite{PH86}, Henrichs
\cite{He88}) showed that these narrow components are variable in
velocity and profile. Continuous time series of UV spectra revealed
that narrow components evolve from broad absorption features, which
appear at low velocity and accelerate to higher wind velocities while
narrowing (Prinja et al.\ \cite{PH87}, Prinja \& Howarth \cite{PH88},
Henrichs et al.\ \cite{HK88}). The current name of these variable
absorption features is {\it discrete absorption components (DACs)},
which are the most prominent features of variability in hot-star
winds.  Because of their specific shape, DACs are readily recognized
in single ``snapshot'' spectra. Howarth \& Prinja (\cite{HP89})
detected DACs in more than 80\% of the IUE spectra in a sample of 203
galactic O~stars. Grady et al.\ (\cite{GB87}) and Henrichs
(\cite{He88}) also found DACs in many Be stars, although not in
non-supergiant B stars. Thus, the occurrence of DACs is a fundamental
property of hot-star winds, and knowing how DACs develop is considered
essential for our understanding of stellar-wind physics.

In search for the origin of stellar-wind variability, several
monitoring campaigns have been organized with IUE. Since the typical
timescale of variability is on the order of hours to days, continuous
monitoring over a period of several days is essential. For this
reason, dozens of NASA and ESA ``shifts'' had to be scheduled in
sequence to the same observing program. On the basis of these
observing campaigns it became clear that wind variability is
systematic. Figures 4 and 5 show examples of the DAC behaviour in the
\siiv\ resonance doublet of the O9.5~Ib supergiant 19~Cep and the
O7.5~III giant 68~Cyg. The bottom panel gives a grey-scale
representation of the evolution of DACs as a function of time (running
upwards). The DACs start at low velocity ($v_{c}=0.2-0.5$~\vinf) as
broad absorption features and, while narrowing, move towards the blue
until they reach an asymptotical velocity. This velocity is
systematically lower (by 10-20\%) than the maximum ``blue edge''
velocity observed in saturated P~Cygni lines. The proposed solution is
that the asymptotic DAC velocity has to be identified with the
``real'' \vinf\ of the stellar wind and that the additional blue-shift
in saturated P~Cygni lines is caused by turbulence (i.e. the local
spread in velocity due to the shocked wind structure, see section
2.2). Thus, DACs provide a diagnostic to measure \vinf; fortunately,
Prinja et al.\ (\cite{PB90}) demonstrated that the so-called $v_{\rm
black}$, the last saturated point in a P~Cygni profile measured in the
direction of increasing blue-shift, shows a tight correlation with the
corresponding asymptotic DAC velocity, so that it is still possible to
obtain \vinf\ from a single UV spectrum.

\begin{figure}[ht]
\centerline{\epsfig{file=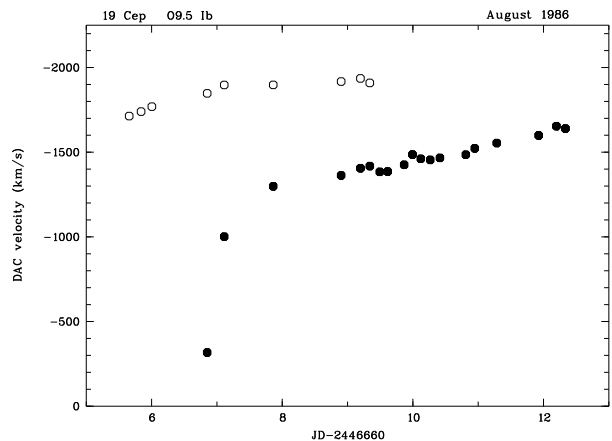,width=8.0cm,angle=-90}}
\caption[]{\em 19~Cep O9.5Ib: The DAC central velocity as a function
of time in the \siiv\ resonance doublet (see Figure~4).}
\end{figure}

Obviously, in saturated resonance lines DACs cannot be observed; the
steep blue edge of these profiles, however, often shows regular shifts
of up to 10\% in velocity, on a timescale comparable to that of the
DACs. This edge variability is presumable related to the DAC
behaviour; the observed differences in the asymptotic velocities of
the DACs combined with optical depth effects might cause the less
regular behaviour of the high-velocity edges (cf.\ Kaper et
al. \cite{KH98}).

\begin{figure}[ht]
\centerline{\epsfig{file=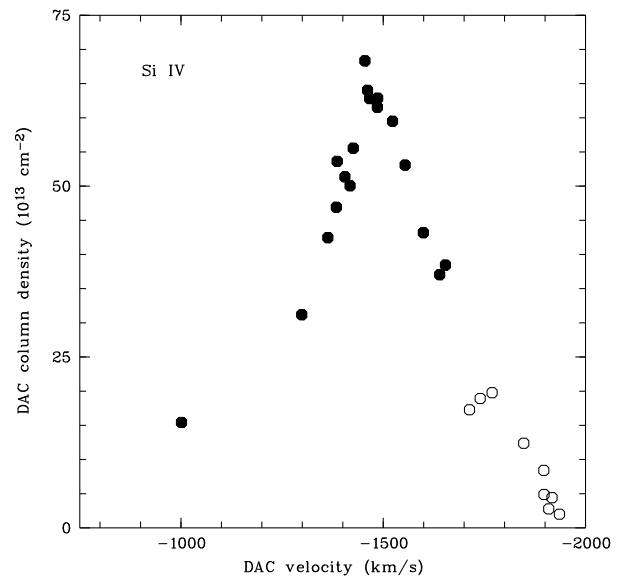,width=8.0cm,angle=-90}}
\caption[]{\em 19~Cep O9.5Ib: The corresponding change in DAC column
density as a function of DAC central velocity in the \siiv\ resonance
doublet. The column density peaks at 0.75~\vinf.}
\end{figure}

Figures 6 and 7 show the results of a quantitative analysis of the DAC
behaviour in the \siiv\ resonance lines of 19~Cep (see also Figure~4).
The migrating DACs were modelled in the way described by Henrichs et
al. (\cite{HH83}) after division of the spectrum by a
minimum-absorption template (Kaper et al.\ \cite{KH98}). The central
velocity, central optical depth, width, and column density was
measured for each pair of DACs in the UV resonance doublets. In
Figure~6 a newly formed DAC accelerates towards the asymptotic
velocity already reached by the narrow absorption component present
from the beginning of the observing campaign. The acceleration is much
slower than expected for a ``normal'' $\beta=1$ velocity law; the
acceleration timescale appears to be longer for stars with a longer
DAC recurrence timescale (see below). The corresponding change in
column density is displayed in Figure~7. As is also found to be true
in general, the DACs reach a maximum $N_{\rm col}$ at a velocity of
about 0.75~\vinf\ and subsequently fade in strength. In cases where
more than one resonance doublet was available to measure DACs,
consistent results were obtained, supporting the interpretation that
DACs correspond to changes in wind density and/or velocity rather than
changes in the ionization structure of the stellar wind. It turns out
that the DAC ``pattern'' is characteristic for a given star, but
differences are found from year to year (e.g. the strength of the
individual DACs, see Kaper et al.\ \cite{KH96}, \cite{KH98}).

\begin{figure}[ht]
\centerline{\epsfig{file=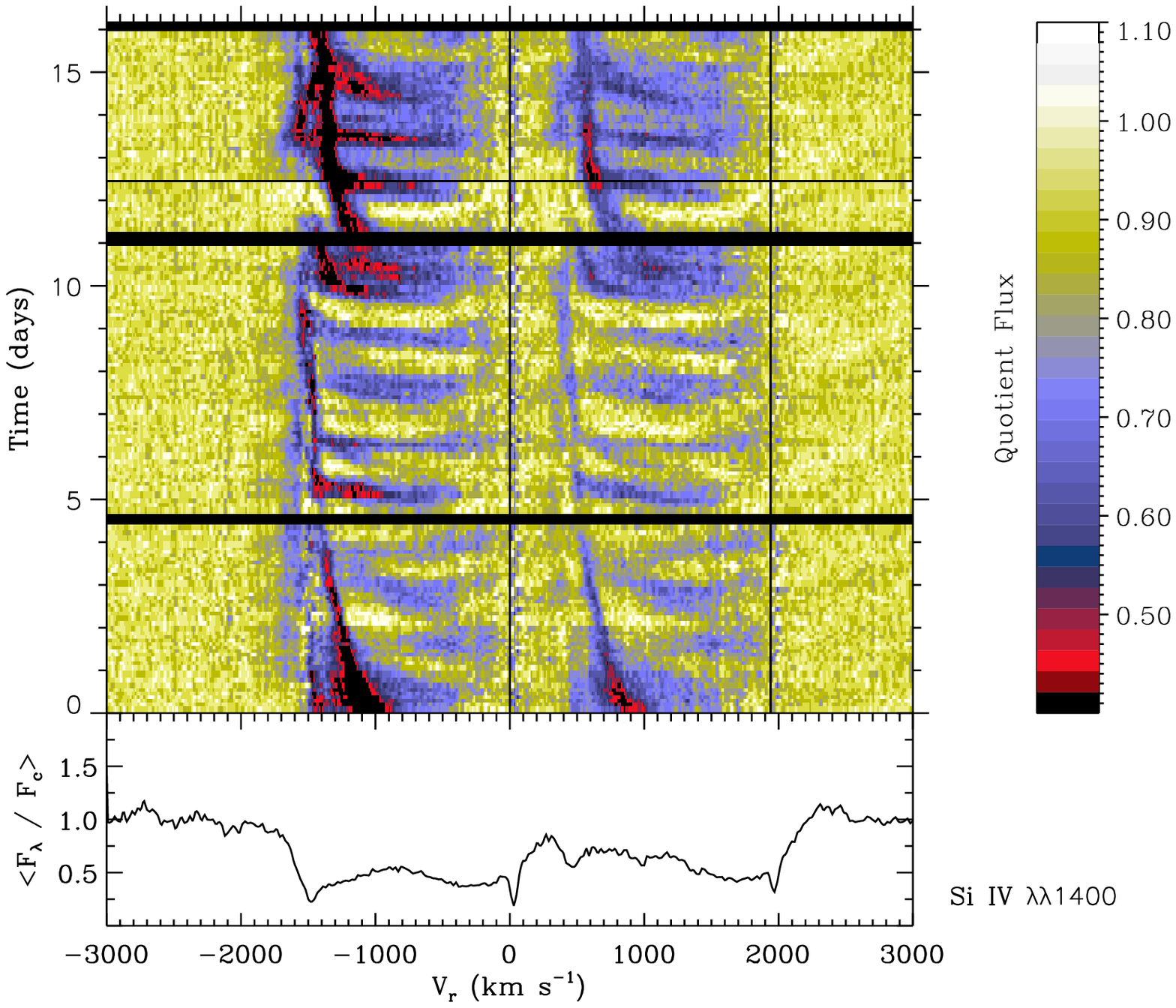,width=8.0cm}}
\caption[]{\em Time series of the \siiv\ resonance doublet of HD64760
($v \sin{i} =$238~km/s) obtained in January 1995 during the IUE MEGA
campaign. Time (in days from the beginning of the series) increases
upward. The spectra were converted to a linear time grid by
interpolation, and gaps appear whenever more than 5 hours elapsed
between exposures. The individual spectra are normalized by a minimum
absorption template so that all changes appear as absorption. The rest
wavelengths of the doublet are shown as vertical lines. Note the
apparent bowing of the black and white stretches (figure taken from
Massa et al.\ \cite{MF95}).}
\end{figure}

\begin{table*}[t]
\begin{center}
\caption[]{\em Measured ``wind'' period in O~star spectra compared to
the estimated maximum rotation period (based on the stellar parameters
listed by Howarth \& Prinja (\cite{HP89}) and references
therein). Runaway stars are indicated by (R). References: 1) Kaper
\cite{Ka93}; 2) Prinja et al.\ \cite{PB92}; 3) Howarth et al.\
\cite{HB93}, \cite{HP95}; 4) Fullerton et al.\ \cite{FG92}: from
He~{\sc i} 5876~\AA; 5) Prinja \cite{Pr88}; 6) Stahl et
al. \cite{SK96}; 7) Kaper et al.\ \cite{KH97}; 8) Kaper et
al. \cite{KF98}.}  
\leavevmode
\begin{tabular}{llcccccc} \hline
Name & Sp.\ Type & $v\sin{i}$ & $R$ & $P_{{\rm max}}$ & 
\multicolumn{2}{c}{$P_{{\rm wind}}$ (days)} & Reference \\ 
 & (Walborn) & (\kms) & ($R_{\odot}$) & (days) & UV & H$\alpha$ & \\ \hline
$\zeta$~Oph (R)  & O9.5 V          & 351 & \08 &\01.2 & 0.9 & & 3 \\
68 Cyg (R)        & O7.5 III:n((f)) & 274 &  14 &\02.6 & 1.4 &
1.3 & 1,7 \\
$\xi$ Per (R)     & O7.5 III(n)((f))& 200 &  11 &\02.8 & 2.0 &
2.0 & 1,7 \\
$\lambda$ Cep (R) & O6 I(n)fp       & 214 &  19 &\04.5 & 1.3 & 
1.2 & 1,7 \\
$\zeta$~Pup (R)         & O4 I(n)f        & 208 &  19 &\04.6 & 0.8 &
0.9 & 2,8 \\
HD34656                 & O7 II(f)        & 106 &  10 &\04.8 & 1.1 & & 1 \\
15 Mon                  & O7 V((f))       & \063 &  10 &\08.0 &$>4.5$ &
& 1 \\
63 Oph                  & O7.5 II((f))    &\080 &  16 & 10.2 &$>3$ & & 5 \\ 
HD135591                & O7.5 III((f))   &\065 &  14 & 10.9 & & 3.1 &
8 \\
$\lambda$ Ori A         & O8 III((f))     &\053 &  12 & 11.5 &$\sim 4$
& 2.0: & 1 \\
$\theta^{1}$ Ori C      & O6-O4 var       & 50: & 12: & 12: &
``15.4'' & 15.4 & 6 \\
19 Cep                  & O9.5 Ib         &\075 &  18 & 12.1 & 4.5 &$\sim 5$
& 1,7 \\
$\mu$ Nor               & O9.7 Iab        &\085 & 21 & 12.5 & & 6.0 &
8 \\
$\alpha$ Cam (R)  & O9.5 Ia         &\085 &  22 & 13.2 & & 5.6 &
8 \\
$\zeta$ Ori A           & O9.7 Ib         & 110 &  31 & 14.3 &$\sim 6$
& 6 & 1,8 \\
10 Lac                  & O9 V            &\032 & \09 & 15.3 &$\sim 7$
& & 1 \\ 
HD112244 (R)      & O8.5 Ib(f)      &\070 &  26 & 18.8 & & 6.2 &
8 \\
HD57682 (R)       & O9 IV           &\017 &  10 & 29.8 & & $>6$
& 8 \\  
HD151804                & O8 Iaf          & \050 &  35 & 35.4 & & 
2-3,7.3 & 4,8 \\
\hline
\end{tabular}
\end{center}
\end{table*}

A key issue is the recurrence timescale of DACs; starting with
Henrichs et al.\ (\cite{HK88}) and Prinja (\cite{Pr88}), all papers in
which more than one sequence of DACs is described, DACs repeat on a
timescale comparable to the estimated rotation period of the star
(cf.\ Kaper et al.\ \cite{KH96}). As is illustrated by Figures 4 and 5,
the star with the higher value of projected rotational velocity $v
\sin{i}$ (68~Cyg, $v \sin{i}=274$~km/s) shows DACs appearing with a
much higher frequency than 19~Cep ($v \sin{i}=75$~km/s). Fourier
analyses performed on the obtained datasets clearly reveal this
periodicity (see, e.g., Kaper et al.\ \cite{KH98}). In Table~1 the
observed ``wind periods'' (i.e., DAC recurrence timescales) are listed
for a sample of O-type stars. The stars are ordered according to the
estimated maximum rotation period:
\[ P_{\rm max} = 50.6 (v \sin{i})^{-1} \left(
\frac{R_{\star}}{R_{\odot}} \right) {\rm days}, \]
where $R_{\star}$ is the stellar radius. The table shows that stars
with a shorter wind period end up higher on the list, consistent with
the interpretation that the wind period relates to the rotation period
of the star.

This strongly suggests that the regular appearance of DACs is related
to the rotation of the star. The nature of this relationship has
recently been the focus of a number of large IUE observing
campaigns. During the IUE MEGA campaign (Massa et al.\ \cite{MF95}) the
WN5 star HD50896 (EZ~CMa), the B0.5~Ib star HD64760, and the O4~If(n)
star HD66811 ($\zeta$~Pup) were monitored for a period of 16 days of
nearly continuous observations in January 1995. The obtained time
series for the \siiv\ resonance lines of HD64760 ($v
\sin{i}=238$~km/s, $P_{\rm max} = 4.8$~days) is shown in Figure~8. The
three stars show continuous wind activity throughout the run with an
apparent periodicity, which is coherent for at least a few rotation
cycles (though not necessarily on longer intervals). Although the
cyclical behaviour of wind variability is confirmed for the three
target stars, the observations raise some important questions. In the
case of $\zeta$~Pup and HD64760, the regular pattern of DACs
(recurrence timescale 0.8 and 1.2 days, respectively) seems to be
superposed on a ``slower'' pattern of additional absorption. For
$\zeta$~Pup the slow pattern gives rise to a period of 5.2~days, which
is interpreted as the rotation period of the star (Howarth et
al. \cite{HP95}). In HD64760 such a ``slow'' period is not found. In
fact, for HD64760 it turned out that the fast pattern is caused by
almost sinusoidal modulations of the flux in stead of pure-absorption
events. The features that we are used to call DACs make up the slow
pattern, but a relation with the fast pattern in HD64760 is not
evident (Prinja et al.\ \cite{PM95}, Fullerton et al.\ \cite{FM97}).

\begin{figure}[ht]
\centerline{\epsfig{file=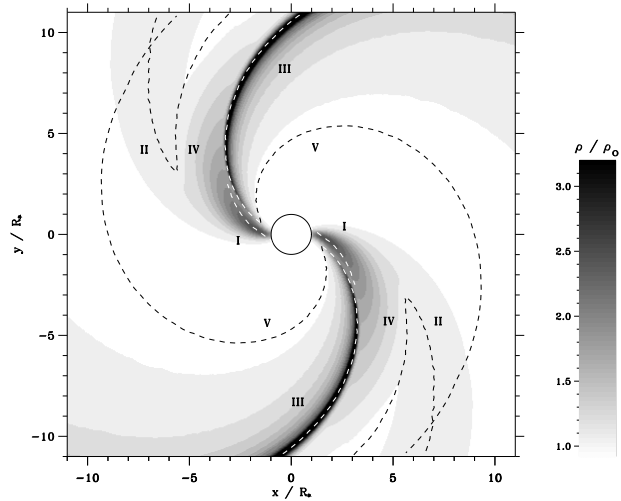,width=8.0cm,angle=90}}
\caption[]{\em Example of the large-scale wind structure produced by
the CIR model of Cranmer \& Owocki. The density structure of the
stellar wind in the equatorial plane (normalized to the
``unperturbed'' wind density) is shown for the case of the Bright Spot
model. Area I indicates the enhanced mass flux above the bright spot,
area III is the CIR compression. According to their model, the largest
relative contribution to the Sobolev optical depth (producing DACs in
line profiles) comes from region V, the so-called radiative acoustic
Abbott kink (figure from Cranmer \& Owocki (\cite{CO96}).}
\end{figure}

A new effect discovered in the spectra of HD64760 is the so-called
phase bowing. Since the Fourier analysis is carried out for each
wavelength point, the phase of the periodic variation is known as a
function of velocity. A given phase of the modulation is first reached
at an intermediate velocity of about 750~km/s and subsequently at both
lower and higher velocities. Close inspection of Figure~8 reveals this
effect as well. Inspired by Mullan (\cite{Mu86}), Owocki et
al. (\cite{OC95}) proposed that these characteristics naturally arise
from absorption by strictly accelerating corotating wind streams seen
in projection against the stellar disk. Cranmer \& Owocki
(\cite{CO96}) worked this out in more detail on the basis of
hydrodynamical simulations. The Corotating Interaction Region (CIR)
model, which was first applied to the solar wind, invokes fast and
slow wind streams that originate at different locations on the stellar
surface. Due to the rotation of the star, the wind streams are curved,
so that fast wind material catches up with slow material in front,
forming a shock at the interaction region. The shock ``pattern'' in
the wind is determined by the boundary conditions at the base of the
wind and corotates with the star. The wind material itself flows in a
(nearly) radial direction under conservation of its angular momentum,
but does {\it not} corotate with the star. It meets the interaction
region at a distance from the star depending on a variety of
parameters, including its original location on the stellar
surface. Cranmer \& Owocki induced an azimuthal variation in the
outflow by a local increase or decrease in the radiative driving
force, as would arise from a bright or dark ``star spot'' in the
equatorial plane. Above a bright spot the mass-loss rate is enhanced
and the corresponding wind stream will reach a lower terminal
velocity. The resulting wind structure is shown in Figure~9. Every
time a CIR passes through the line of sight, a DAC is observed,
explaining both the relation between the DAC recurrence timescale and
stellar rotation and the observed phase bowing.

Also the \Ha\ profiles of early-type stars indicate dramatic
variability (Ebbets \cite{Eb82}). The origin of this variability has
to be understood given the importance of this line for mass-loss
determinations (section 2.1). Kaper et al.\ (\cite{KH97}) show that
for their sample of O-type stars, simultaneous ultraviolet and \Ha\
observations reveal the same periodicity and are consistent with the
CIR model. Thus, wind variability also affects the \Ha\ line, although
for main sequence stars the line is too weak to detect variability. In
Table~1 a comparison is made between the wind periods derived from the
UV P~Cygni lines and the \Ha\ profile. The results of a search for
wind variability in the \Ha\ line for a large sample ($\sim 70$) of
O-type stars will be presented in Kaper et al.\ (\cite{KF98}).

In conclusion, the CIR model seems to be the best explanation for the
observed wind variability. The origin of the bright or dark spots on
the stellar surface is, however, not known. The two obvious candidates
are non-radial pulsations and surface magnetic fields. Reference is
made to the contribution by Henrichs in this volume for a more
detailed interpretation of the cyclical variability in hot-star winds.

\section*{ACKNOWLEDGEMENTS}

I would like to thank my colleagues Huib Henrichs, Joy Nichols, Jeroen
de Jong, Alex Fullerton, Joachim Puls, Stan Owocki, and John Telting
for the nice collaboration. The IUE Observatory staff at both Goddard
and Vilspa are acknowledged for their dedicated efforts in executing
these difficult programs. I thank the Organizing Committee for the
invitation to present this paper.

\end{document}